\documentclass[12pt,preprint]{aastex}
\usepackage{emulateapj5}

\shorttitle{Obscuring Clouds}
\shortauthors{Watabe and Umemura }

\begin{document}

\title{Obscuration of Active Galactic Nuclei by
Circumnuclear Starbursts}

\author{YASUYUKI WATABE AND MASAYUKI UMEMURA}
\affil{Center for Computational Sciences, University of
Tsukuba, Ten-nodai, 1-1-1 Tsukuba, Ibaraki 305-8577,
Japan; watabe@rccp.tsukuba.ac.jp, umemura@rccp.tsukuba.ac.jp}

\begin{abstract}

We examine the possibility of the active galactic nucleus (AGN) obscuration
by dusty gas clouds that spurt out from circumnuclear starburst regions.
For the purpose, the dynamical evolution of gas clouds is pursued, 
including the effects of radiation forces 
by an AGN as well as a starburst. 
Here, we solve the radiative transfer equations for clouds, 
taking into consideration the growth of clouds by
inelastic cloud-cloud collisions 
and the resultant change in optical depth. 
As a result, it is shown that if the starburst is more luminous
than the AGN, gas clouds are distributed extensively
above a galactic disk with the assistance of radiation pressure 
from the starburst. The total covering factor of gas clouds 
reaches a maximum of around 20\%. After several $10^{7}$yr,
gas clouds with larger optical depth form by cloud-cloud
collisions and thereafter the clouds fall back due to
weakened radiation pressure. The larger clouds undergo
runaway growth and are eventually distributed around the
equatorial plane on the inner sides of circumnuclear
starburst regions. These clouds have an optical depth of
several tens. 
The result is qualitatively consistent with the putative tendency that 
Seyfert 2 galaxies appear more frequently associated with starbursts
than Seyfert 1s. 
On the other hand, if the AGN
luminosity overwhelms that of the starburst,
almost all clouds are ejected from the galaxy due to the radiation pressure 
from the AGN, resulting in the formation of a quasar-like object.
The origin of obscuration of AGNs is discussed with 
relevant observations.
\end{abstract}


\section{INTRODUCTION}

The origin of the active galactic nucleus (AGN) obscuration is a key issue
for a thorough understanding of AGNs. 
In the context of the unified model, 
an obscuring torus has been thought to be responsible for the
AGN obscuration, and the dichotomy of AGN types has been attributed to
the viewing angle toward the AGN 
 (see Antonucci 1993 for a review). 
However, recent observations on circumnuclear regions of AGNs 
have gradually revealed that 
Seyfert 2 galaxies appear to be more frequently
associated with starbursts than Seyfert 1s
(Heckman et al. 1989; Maiolino et al. 1997,
1998a; Perez-Olea \& Colina 1996; Hunt et al. 1997;
Malkan, Gorjian, \& Tam 1998;
Schmitt, Storchi-Bergmann, \& Fernandes 1999; 
Storchi-Bergmann et al. 2000;
Gonzalez Delgado, Heckman, \& Leithere 2001). 
Furthermore, quasars are mostly observed as
type 1 AGNs, regardless of the star formation activity
in host galaxies (Barvainis, Antonucci,
\& Coleman 1992; Ohta et al. 1996; Omont et al. 1996;
Schinnerer, Eckart, \& Tacconi 1998; Brotherton et
al. 1999; Canalizo \& Stockton 2000a, 2000b; Dietrich \&
Wilhelm-Erkens 2000; Solomon et al. 2004). 
In addition, Ueda et al. (2003) have recently 
found from the hard X-ray luminosity function of AGNs that 
the fraction of X-ray absorbed AGNs decreases with the AGN luminosity.
These recent findings seem beyond
understanding in the context the unified model.

The frequent association of circumnuclear starbursts with
Seyfert 2 galaxies suggests that the obscuring of AGNs 
could be physically related to the starburst events.
Ohsuga \& Umemura (1999, 2001) have considered a physical mechanism
to link the AGN type to the circumnuclear starburst, where
an obscuring wall composed of dusty gas can form 
with the assistance of the radiation pressure from circumnuclear starbursts.
More recently, Wada \& Norman (2002) have performed
three-dimensional hydrodynamic simulations
on the nuclear starburst regions, and found that 
the gas blown out from a galactic disk
due to multiple supernova explosions in a starburst
builds up a torus-like structure 
in a highly inhomogeneous and turbulent manner.
Both models may provide potential mechanisms to relate
the starburst events with the obscuring of AGNs.
But, Ohsuga \& Umemura focused just on the
equilibrium configuration of dusty gas,
and Wada \& Norman did not include
radiation forces, which are likely to play 
an important role in the early phase of starbursts. 
Hence, in order to construct a more realistic model, 
the dynamical evolution of obscuring materials 
should be solved, including radiation forces.
This requires a radiation-hydrodynamic (RHD) simulation 
in three-dimensional space.
However, a full RHD simulation 
is extremely time consuming even using the state-of-the-art 
computational facilities. 
In this paper, we solve the dynamics of discrete clouds 
under the adhesion approximation, where the growth of
clouds is brought by inelastic cloud-cloud collisions.
Simultaneously, to properly include the radiation forces,
we solve the radiation transfer equation for clouds,
taking the change in optical depth due to the cloud growth into account.
In \S 2, the circumnuclear regions and gas clouds are modeled.
In \S 3, 
we formulate the equation of motion for gas clouds
including radiation forces. In \S 4,
the numerical results are shown, and the
distribution of gas clouds and the obscuration of AGN
are analyzed. In \S 5, based on the present results, 
we discuss the origin of AGN obscuration and the relation to 
narrow line regions.
\S 6 is devoted to the conclusions.

\section{MODEL}

Recent high-resolution observations 
have revealed circumnuclear starburst regions
with a radial extension of 10 pc up to 1 kpc, 
which frequently exhibit a ring-like feature 
(Pogge 1989; Wilson et al. 1991; Forbes et al. 1994; Marconi
et al. 1994; Mauder et al. 1994; Buta, Purcell, \&
Crocker 1995; Barth et al. 1995; Leitherer et al. 1996;
Maoz et al. 1996; Storchi-Bergmann, Wilson, \& Baldmin
1996; Elmouttie et al. 1998). 
In this paper, we suppose a starburst region in ring configuration 
with a radius of 200 pc and the total mass of starburst
ring is assumed to be $M_{\rm{SB}} = 10^{8}M_{\odot}$,
similar to the model by Ohsuga \& Umemura (2001).
Also, we settle a galactic bulge component with
the mass of $M_{\rm{GB}} = 10^{10}M_{\odot}$
and the radius of $R_{\rm{GB}} = 1$ kpc.

In some galaxies, a galactic bulge solely cannot explain
the observed rotation curve. For instance, in the circumnuclear
regions of the Circinus galaxy, the rotation velocity 
requires an additional component with
$10^{9-10}M_{\odot}$ within a starburst ring
(Elmouttie et al. 1998). Also, the stellar velocity dispersion 
suggests that the Circinus galaxy has a spread 
component within a radius of several 100 pc 
(Maiolino et al. 1998a). In this paper, with taking
such mass distributions into consideration, 
we also assume an inner bulge component
with the mass of $M_{\rm{IB}} = 10^{9}M_{\odot}$
and the radius of $R_{\rm{IB}} = 100$ pc.
As for a central supermassive black hole (SMBH),
the mass is assessed in terms of the recently inferred
black hole-to-bulge mass relation, that is,
$M_{\rm BH}/M_{\rm bulge} \approx 10^{-3}$
 (Richstone et al. 1998; Marconi \& Hunt 2003, and references therein). 
Here, the SMBH mass is set to be $M_{\rm BH} =10^{7}M_{\odot}$.

We assume the energy spectrum of AGN to be in the form of
$L_{\rm{AGN}}^{\rm{\nu}} \propto \nu^{-1}$ between 0.01
and 100 keV (Blandford et al. 1990). 
The bolometric luminosity, $L_{\rm AGN}$, is set
to be constant for a period of $10^{8}$ yr,
which is a typical age of AGN (basically Eddington timescale). 
In the starburst regions, we assume a Salpeter-type 
stellar initial mass function (IMF) 
for a mass range of [$m_{\rm l}, m_{\rm u}$];
\begin{equation}
\phi (m_{\ast}) = \phi_{0} 
(m_{\ast}/M_{\odot})^{-(1+s)},
\end{equation}
where $s = 1.35$ and $m_{\ast}$ is the stellar mass.
Although $m_{\rm l}$ and $m_{\rm u}$ in starburst regions are
under debate, some authors argue that the IMF in a starburst 
is deficient in low-mass stars, where
$m_{\rm{l}} \approx 2M_{\odot}$ is derived. 
The upper limit is inferred to be $m_{\rm{u}} \approx
40M_{\odot}$ (Doyon, Puxley, \& Joseph 1992; Charlot et
al. 1993; Doane \& Mathews 1993; Hill et al. 1994;
Blandle et al. 1996).
We assume these values for $m_{\rm l}$ and $m_{\rm u}$.
The star formation rate, $\dot{M}_{\ast}$,
is controlled by $\dot{M}_{\ast} =
(M_{\rm SB}/t_{\rm SF})
\exp(-t/t_{\rm SF})$, where $t$ is the elapsed
time after the initial starburst and $t_{\rm SF}$ is
the duration of star formation, which is
assumed to be $10^{7}$yr (e.g. Efstathiou, Rowan-Robinson, \& 
Siebenmorgen 2000). 
We employ a stellar mass-to-luminosity relation as
$(l_{\ast}/L_{\odot}) =(m_{\ast}/M_{\odot})^q$ 
with $q = 3.7$, and
a stellar mass-to-age relation as 
$t_{\ast} =1.1\times 10^{10}{\rm yr} (m_{\ast}/M_{\odot})^{\omega} $ 
with $\omega = 2.7$ (Lang 1974). 
All stars are assumed to emit the radiation in the blackbody.

\begin{figure}[hbtp]
  \centering
	\includegraphics[width=10cm]{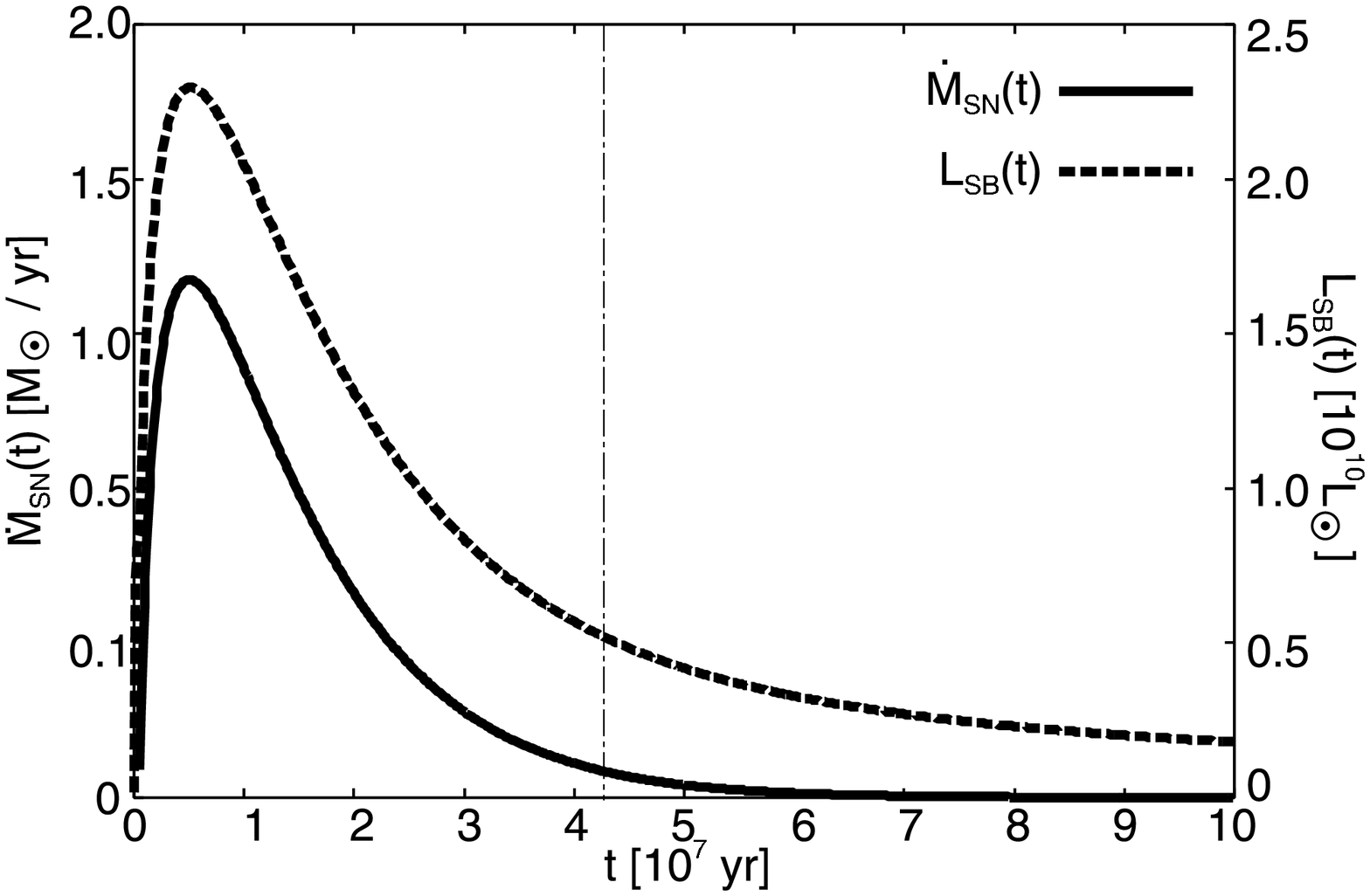}
	\caption{The evolution of mass ejection rate
	({\it solid line}) and luminosity of starburst region
	({\it dashed line}). The starburst is super-Eddington
	luminous for gas clouds of initial
	size before the epoch indicated by a dot-dashed line
 ($\sim 4.3\times 10^{7}$ yr).}
	\label{fig1}
\end{figure}

Next, we model the mass ejection from the starburst
regions. The mass ejection is driven by multiple supernova (SN)
explosions in a starburst. 
Stars heavier than $m_{\rm{crit}} = 8M_{\odot}$ are 
destined to undergo Type II SN explosions.
The SNe restore almost all the mass into interstellar space and
release the energy with a conversion efficiency to the rest mass energy as
$\varepsilon \approx 10^{-4}$.
Thus, the rate of mass restoration by SNe is assessed by
$\dot{M}_{\rm{SN}}(t) = L_{\rm{SN}}(t)/ c^{2}\varepsilon$, 
where $M_{\rm{SN}}(t)$ is the total mass of
SNe that explode by time $t$ and
$L_{\rm{SN}}(t)$ is the total supernova
luminosity that is temporally averaged.
The energy input by multiple SN explosions 
leads to the formation of 
an elongated hot cavity ({\it superbubble}), and consequently
the shock-heated interstellar gas spurts out from the 
starburst regions
(Shapiro \& Field 1976; Tomisaka \& Ikeuchi 1986; 
Norman \& Ikeuchi 1989). 
A recent simulation by Wada \& Norman (2002) have shown that
the shock-heated gas eventually fragments into 
cold, dense gas clouds owing to the cooling. 
Hence, we simply suppose that gas clouds spurt from
the starburst regions with the mass restoration rate
estimated above. 
(It is noted that a part of preexisting interstellar gas could be 
also accelerated by the superbubble and therefore participate in the
mass ejection from a starburst. Hence, the present mass ejection
rate is regarded as a minimum rate.) 
On the basis of Wada \& Norman's results, we set the initial cloud
radius and mass to be $r_{\rm{c}} = 3.5$ pc and
$M_{\rm{c}} = 10^{3}M_{\odot}$ as a fiducial model. Then, 
the hydrogen density in a cloud is 
$2\times 10^2~{\rm cm}^{-3}$.
We also examine the effect of a different initial size of clouds
by setting $r_{\rm{c}} = 2.5$ pc or $r_{\rm{c}} = 4.5$ pc.
The cloud velocity is assumed to be $V_{\rm{c}} \sim 150$ km/s,
which is on the order of escape velocity. 
Figure \ref{fig1} shows the evolution of mass ejection
rate (solid line) and luminosity (dashed line) 
in the starburst regions. Before $4.3\times 10^{7}$ yr,
the starburst is super-Eddington
luminous for gas clouds of initial size. 
As for the effects of photoionization and photodissociation
by the radiation from a starburst as well as an AGN,
Ohsuga \& Umemura (2001) have analyzed the inner structure
of an optically-thick gas slab by 
solving the force balance and the energy equation.
They found that
the dust cooling is still effective for a gas slab located
at several hundred parsecs from the galactic center, and 
the resultant gas temperature becomes on the order of 100 K. 
Hence, we assume the cloud temperature to be 100 K and
the optical depths of clouds to be determined by the dust opacity.

The spurted clouds can collide with each other in the space
above the galactic disk. 
Here, we simply assume that an inelastic 
collision occurs if the distance between cloud
centers is less than the sum of cloud radii. 
Also, it is assumed 
that the density of clouds does not alter after the collision, so
that a cloud of larger size is generated by the collision. 
Taking into account the resultant change in optical depth of clouds,
we calculate the dilution of the radiation from a starburst and an AGN.
It should be noted that this assumption on the collision is quite naive. 
Nagasawa \& Miyama (1987) have simulated the head-on collisions
of isothermal clouds, using a 3D hydrodynamic code, and
have found that the cloud-cloud collisions are very inelastic 
even for the Mach number of 10, and they result
in the coalescence without disruption.
However, in our simulation, the Mach number 
can be up to several 10 or $\sim$ 100.
Also, the off-center collisions can frequently occur. 
In this case, 
the cloud-cloud collision may result in not only the
coalescence but also the disruption of clouds or the induction of star
formation. 
These effects seem worth incorporating in the future 
simulations. 

The self-gravity of 
clouds is $\sim 10^{-3}$ of the external gravity
in the model galaxy.
Hence, the self-gravity of clouds can be neglected.
But, the growth of clouds induced by collisions can lead to
the gravitational instability, if the cloud radii exceed the
Jeans scale, which is 10.8 pc for the
assumed cloud temperature of 100 K.
In the present simulation, we treat a cloud above
the Jeans scale as a collapsing cloud which 
contributes no more to the obscuration.

\section{BASIC EQUATIONS}

The equation of motion is given by 
\begin{equation}
\frac{d^{2}\vec{r}}{dt^{2}} = \vec{f}_{\rm{rad, SB}} +
\vec{f}_{\rm{rad, AGN}} + \vec{f}_{\rm{grav, SB}} +
\vec{f}_{\rm{grav, BH}} + \vec{f}_{\rm{grav, bulge}},
\end{equation}
where $\vec{r}$ is the position vector of a cloud
from the galactic center, and 
$\vec{f}_{\rm{grav, SB}}$, $\vec{f}_{\rm{grav, BH}}$, and
$\vec{f}_{\rm{grav, bulge}}$ are 
the gravitational forces from the starburst
ring, the central SMBH, and the bulges, respectively.
The radiation pressure force from the starburst, $\vec{f}_{\rm{rad, SB}}$, 
is given by 
\begin{equation}
\vec{f}_{\rm{rad, SB}} = \int_{\rm{V}}dV
\frac{\bar{\chi}_{\rm{SB}}}{c}\frac{\rho_{\rm{SB}}(t)}{4\pi
  |\vec{l}_{\rm{SB}}|^{3} }\vec{l}_{\rm{SB}}\left[ \frac{1-\exp
(-\tau_{\rm{SB}})}{\tau_{\rm{SB}}} \right], 
\end{equation}
where $\bar{\chi}_{\rm{SB}}$ is the mass extinction coefficient, 
which is averaged over the spectrum of the starburst;
$\bar{\chi}_{\rm{SB}} =
\int_{\rm{\nu}}\chi_{\rm{\nu}}L_{\rm{SB}}^{\rm{\nu}}d\nu/L_{\rm{SB}}$,
where $\chi_{\rm{\nu}}$ is the mass extinction coefficient 
and $L_{\rm{SB}}^{\rm{\nu}}$ is the spectrum. 
Since the mass density is dominated by gas
with little contribution of dust and the 
Thomson scattering is negligible in the opacity
(Umemura, Fukue, \& Mineshige 1997, 1998), 
$\chi_{\rm{\nu}}$ is given by $\chi_{\rm{\nu}} =
\kappa_{\rm{\nu}}^{\rm{d}}/\rho_{\rm{g}}$, where
$\kappa_{\rm{\nu}}^{\rm{d}}$ is the opacity by dust grains 
per unit volume and $\rho_{\rm{g}}$ is the gas
density in a cloud. 
$\tau_{\rm{SB}}$ is the cloud optical depth 
estimated by $\bar{\chi}_{\rm{SB}}$,
$\rho_{\rm{SB}}(t)$ is the luminosity density in the
starburst region, and $\vec{l}_{\rm{SB}}$ is the position vector 
of a cloud from a volume element $dV$ in the starburst region. 
We suppose the dust-to-gas mass ratio
to be $0.03$, which is three times as high as that observed
in the solar neighborhood, because
the metallicity in QSOs is found to be several 
to ten times higher than the solar metallicity
(e.g. Hamann \& Ferland 1993).  
Here, we employ
a grain size distribution of a power-law as
$n_{\rm{d}}(a_{\rm{d}}) \varpropto
a_{\rm{d}}^{-3.5}$ in the range of 0.01-1 $\rm{\mu m}$,
which is found in the interstellar matter
(Mathis, Rumpl, \& Nordsieck 1977), and the
absorption cross section is determined by $\pi a_{\rm{d}}^{2}
\min[1,({2\pi \nu a_{\rm{d}}}/c)^{2}]$, where
$a_{\rm{d}}$ is the grain radius. The
density of solid material within a grain is assumed to
be 1.0 g $\rm{cm}^{-3}$ (e.g. Spitzer 1978). 
We find the resultant optical depth of a gas cloud 
to be $\tau_{\rm{c}} \sim 4$.

Similarly to the starburst case, 
the radiation pressure force from the AGN,
$\vec{f}_{\rm{rad, AGN}}$, is calculated by
\begin{equation}
\vec{f}_{\rm{rad, AGN}} =
\frac{\bar{\chi}_{\rm{AGN}}}{c}\frac{L_{\rm{AGN}}}{4\pi
\left| \vec{r} \right|^{3}} \vec{r} \left[
\frac{1-\exp(-\tau_{\rm{AGN}})}{\tau_{\rm{AGN}}} \right],
\end{equation}
where $\bar{\chi}_{\rm{AGN}}$ is the mass extinction coefficient, 
which is averaged over the spectrum of the AGN,
$\tau_{\rm{AGN}}$ is the optical
depth estimated by $\bar{\chi}_{\rm{AGN}}$. 
Here, we neglect the radiation force from SNe,
because the average luminosity of SNe in the assumed IMF 
is less than 10$\%$ of the total stellar luminosity of 
starbursts.

\section{NUMERICAL RESULTS}

Here, we show the numerical results 
for a starburst-dominant case and an AGN-dominant case. 
In each case, the time variations of the 
spatial distributions of gas clouds are presented, and
the obscuration of AGN by the clouds is analyzed.

\subsection{Starburst-Dominant Case}
\begin{figure*}[hbtp]
  \centering
	\includegraphics[width=10cm]{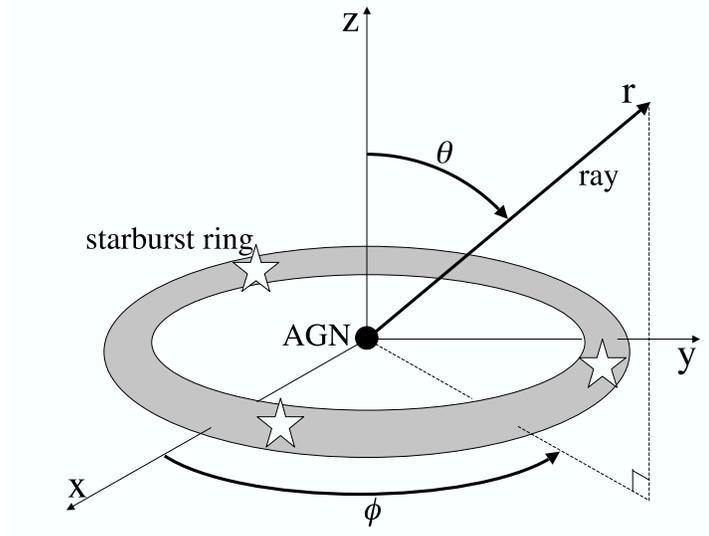}
	\caption{Schematic view of the circumnuclear
	structure and the coordinate system.} 
	\label{fig2}
\end{figure*}

We consider a case with $L_{\rm{AGN}} = 10^{10}L_{\odot}$.
Then the AGN luminosity is lower than the starburst luminosity 
until $\sim 3\times 10^{7}$ yr. 
To show the cloud distributions, 
we use the coordinates depicted in Figure \ref{fig2}.
The resultant cloud distributions projected onto 
the $rz-$ and the $xy-$plane are shown in the left panels in Figure \ref{fig3}.
The colors of points stand for the cloud radii, 
where blue points are 3.5 pc, which corresponds to
the clouds missing collisions, and red ones are the maximum radius
of 10.8 pc, which is the Jeans radius. 
Using the resultant cloud distributions, 
we calculate the optical depth, $\tau$, toward the
direction with $\theta$ and $\phi$. 
Taking the impact parameter into account, $\tau$ is given by 
\begin{equation}
    \tau 
    = \sum_{i,b(i) \leq r_{\rm{c}}(i)}
	2 \left(1-\displaystyle{\frac{b(i)^{2}}{r_{\rm{c}}(i)^{2}}}
	\right)^{\frac{1}{2}}\tau_{\rm{c}}(i), 
\end{equation}
where $i$ denotes each gas cloud, $b(i)$ is the impact parameter 
(the distance from the cloud center to the line-of-sight), 
$\tau_{\rm{c}}(i)$ is the optical depth of the cloud 
along the diameter, and $r_{\rm{c}}(i)$ is the cloud radius. 
If $b(i) \leq r_{\rm{c}}(i)$,
the clouds contribute to the optical depth.
The right panels in Figure \ref{fig3}
show $\tau$ for the each direction
$(\theta, \phi)$, using the Mollweide's projection for the
all-sky plot. The colors represent the levels of $\tau$,
ranging from 0 (white) to 50 (red). 

As shown in Figure \ref{fig3}, 
gas clouds are distributed extensively above the starburst ring
at the stage of $\sim 10^{7}$ yr. 
The radiation pressure from the starburst works to accelerate
clouds, because the starburst is super-Eddington 
luminous until $4.3 \times 10^{7}$ yr for gas clouds of initial size. 
But, clouds are not distributed around the polar regions.
This is because clouds are ejected from the rotating starburst ring
and thus possess angular momenta. 
At this stage, 
the formation of larger clouds by cloud-cloud collisions 
is not promoted to a great extent. 
At $\sim 2\times 10^{7}$ yr, clouds 
with large optical depths increase by repetitive cloud-cloud
collisions and they fall back due to the reduced 
radiation pressure per unit mass. 
The larger clouds undergo 
runaway growth and are eventually distributed around the
equatorial plane on the inner sides of circumnuclear starburst regions. 
These clouds have an optical depth of several tens. 
The angular momentum transfer by cloud collisions near 
the equatorial plane is not so efficient to allow the effective mass 
accretion on to the AGN. Therefore, the large clouds stay there and
can contribute to the obscuration of the AGN. 
At $\sim 10^{8}$ yr,
the starburst luminosity becomes sub-Eddington for all clouds, 
and therefore gas clouds fall back one after another. 
In this stage, a lot of clouds exceed the
Jeans size, and thus the number of obscuring clouds 
decreases rapidly. 
In Figure \ref{fig4}, the transition of the mass spectrum 
of the clouds is shown. 
This figure demonstrates the collision-induced growth of clouds 
and the decrease of obscuring clouds by the Jeans instability. 

Next, we evaluate the covering factor of obscuring clouds. 
Figure \ref{fig5} shows the time variations of the covering factor,
which is defined by the fraction in the all sky of 
the area with the optical depth ($\tau$) larger
than an optical depth given in the abscissa.
If we see the covering factor of $\tau >2-3$, it changes
from $\sim$ 13$\%$ at
$10^{7}$ yr to a maximum of $\sim 21\%$ at $2\times 10^{7}$ yr.
This result shows that the starburst-origin dusty clouds 
can contribute to the obscuration of AGN to a notable extent. 
After that, the covering factor lessens due to the fall back of clouds.

To see the effect by a different choice of the initial size of 
gas clouds, we also calculate the models with 
the initial cloud radius of $r_{\rm{c}} = 2.5$ or 4.5 pc
without changing the cloud mass.
In Figure \ref{fig6}, we show the resultant covering factor 
against the optical depth. 
If the spatial distribution is the same, the shrink of clouds
would result in the reduction of covering factor due to the smaller
geometrical cross section of clouds. 
The reduction of covering factor would be a factor of $(2.5/3.5)^2=0.51$,
if $r_{\rm{c}} = 2.5$ pc is assumed.
But, as seen in Figure \ref{fig6}, the actual reduction of
peak covering factor is $0.16/0.21=0.76$. This can be understood by
the effect of radiation pressure. If the initial cloud size is
smaller, the optical depth of a cloud becomes larger. Hence, 
the radiation pressure on a cloud is weakened, so that clouds
are distributed in more compact space. This effect tends to enlarge 
the covering factor.
On the other hand, for the case of $r_{\rm{c}} = 4.5$ pc, 
the peak covering factor increases by just a small factor 
of $0.23/0.21=1.1$, although
the geometrical cross section would lead to
the increase of covering factor by $(4.5/3.5)^2=1.65$. 
There is again the radiation pressure effect.
Expanded clouds can be distributed more sparsely by less diluted
radiation pressure, so that the covering factor is reduced.
In addition, larger clouds more frequently collide with each other.
This effect also reduces the covering factor. 
As a result, it turns out that the peak covering factor is not
so sensitive to the initial size of clouds.

\subsection{AGN-Dominant Case}

Here, we show the results in an AGN-dominant case, 
where $L_{\rm{AGN}} = 3\times 10^{11}L_{\odot}$
is assumed. In this case, the AGN luminosity is higher than
the starburst luminosity all the time.
Figures \ref{fig7} and \ref{fig8}
show the results by the same quantities as Figures
\ref{fig3} and \ref{fig5}, respectively. 
The assumed AGN luminosity is close to the Eddington
luminosity for the total mass of the galaxy. 
Hence, gas clouds are vigorously accelerated.
Even at the early stage of $\sim 10^{7}$ yr,
clouds are already distributed over kpc, as shown in
Figure \ref{fig7}.
Very few collisions occur, because the 
radiation pressure from the AGN drives a spreading outflow of clouds. 
At $\sim 2\times 10^{7}$ yr, a large fraction of clouds
escape from the galaxy, so that most of gas clouds
cannot contribute to the obscuration of AGN. 
At $\sim 10^{8}$ yr, a small number of 
collision-induced large clouds are left.

Due to the cloud motion driven by the AGN luminosity,
the covering factor of obscuring clouds cannot reach a higher
level. The maximum covering factor is 
$\sim 5\%$ at $10^{7}$ yr, as shown in Figure \ref{fig8}.
Hence, the AGN is likely to be observed as a type 1
with high probability.
Such an AGN-dominant case may correspond to quasar events.
The relation between AGN type and luminosity 
is discussed below in the context of the obscuration. 

\section{DISCUSSION}
\subsection{Origin of Obscuration}

In the previous section, it is shown that the covering factor
by starburst-origin dusty clouds
increases in a luminous phase of the circumnuclear starburst.
This can provide a qualitative explanation for the fact that
Seyfert 2 galaxies are more frequently
associated with starbursts than Seyfert 1s.
On the other hand, obscuring clouds are not distributed around 
polar regions, as shown in Figure \ref{fig3}. 
Thus, in the face-on view, the AGN is likely to be identified as a type 1
even in a luminous phase of starburst, if there is no obscuring materials
other than starburst-origin dusty clouds.
In that sense, Seyfert 1s may be coupled with starburst events. 
Interestingly, hidden starbursts have been found in Seyfert 1 galaxies
(Imanishi \& Dudley 2000; Rodriguez-Ardila \& Viegas 2003), and also
there is an example of Seyfert 1 galaxies accompanied by 
circumnuclear starbursts, that is, NGC 7469 (Genzel et al. 1995). 
In the present simulation, if we see the starburst ring 
at face-on view, starburst-origin dusty clouds
obscure 30\% of the starburst ring at maximum. 
This covering factor is not sufficient to produce perfectly hidden 
starbursts. However, further effects can change the covering factor.
In the present simulation, we have ignored collapsing clouds 
above the Jeans scale. But, the collapsing clouds are likely to 
result in the star formation. Hence,
the re-ejection of matter by supernovae
in the collapsed object may also contribute to the obscuring. 
Furthermore, in the light of the tight correlation
between SMBHs and galactic bulges
 (Richstone et al. 1998; 
Marconi \& Hunt 2003; Kawakatu \& Umemura 2004),
a starburst in a bulge may contribute to the obscuration of nuclear regions
(e.g. Umemura 2001; Kawakatu, Umemura, \& Mori 2003). 

As for the ratio of Seyfert 2s to 1s,
Maiolino \& Rieke (1995) concluded that
Seyfert 2s appear to be four times more numerous than Seyfert 1s.
This implies that the total covering factor of obscuring materials
is likely to be of order 80\%. 
In the present simulation, the maximum covering factor is
around 20\%. 
A hydrodynamic simulation by Wada \& Norman (2002) also shows 
the covering factor to be 40\% at an early phase (1.6Myr) of starburst. 
Hence, it seems that additional obscuring materials are required
to account for the number ratio between Seyfert 1s and 2s.
Recently, we have some significant pieces of observational
information about the obscuring materials. 
The $A_V$ of the nuclear or circumnuclear 
regions is estimated to be between a few and several magnitudes
by IR and optical observations (Rix et al. 1990; Roche
et al. 1991; Goodrich, Veilleux, \& Hill 1994; McLeod \&
Rieke 1995; Oliva, Marconi, \& Moorwood 1999). 
On the other hand, X-ray observations have shown
that most Seyfert 2 nuclei are heavily obscured along
the line of sight with at least $A_V>10$ mag and sometimes 
$A_V>100$ mag
 (Matt et al. 1996, 1999; Maiolino et al. 1998b;
Bassani et al. 1999; Risaliti, Maiolino, \& Salvati 1999). 
Also, it
is argued that a component of obscuring materials must
be extended up to $\geq 100$ pc in addition to a compact component
confined to subparsec scales (Rudy, Cohen, \&
Ake 1988; Miller, Goodrich, \& Mathews 1991; Goodrich
1995; McLeod \& Rieke 1995; Maiolino et al. 1995; Maiolino
\& Rieke 1995; Malkan et al. 1998). These findings
suggest that the distributions of dusty gas around an AGN
are much more diverse than previously considered.
The dusty clouds ejected from starbursts may be 
the origin of the extended obscuring matter.
In addition, inner obscuring materials (possibly a dusty torus) 
may cooperatively work to obscure the nucleus.
The combination of outer and inner obscuring materials
will be an important issue to be scrutinized. 

For the case that the AGN luminosity is predominant, 
dusty clouds are blown out from a galaxy 
by the AGN radiation pressure, and therefore
cannot contribute much to the obscuration.
This result matches the fact that quasars are mostly observed as
type 1 AGNs, regardless of the star formation activity
in host galaxies (Barvainis, Antonucci,
\& Coleman 1992; Ohta et al. 1996; Omont et al. 1996;
Schinnerer, Eckart, \& Tacconi 1998; Brotherton et
al. 1999; Canalizo \& Stockton 2000a, 2000b; Dietrich \&
Wilhelm-Erkens 2000; Solomon et al. 2004). 
Also, based on the hard X-ray 
luminosity function of AGNs, Ueda et al. (2003) have found
that the fraction of X-ray absorbed AGNs 
decreases with the AGN luminosity. This trend is also consistent
with the present picture of the radiation pressure-induced outflow.

\subsection{Relation to Narrow Line Regions}

It may be intriguing to consider the possible 
relation between dusty clouds and narrow line regions (NLRs) 
in AGNs, because the velocity dispersion of 
starburst-origin clouds, a few 100 km s$^{-1}$, 
is just on the order of that found in NLRs. 

One of common properties of NLRs is a fairly constant
ionization parameter, i.e., 
$U\equiv S/nc \approx 0.01$, where $S$ is the ionizing photon flux
and $n$ is the density at irradiated cloud surfaces.
Recently, Dopita et al. (2002) argued that
the ionization parameter of NLRs is successfully accounted for by 
radiation pressure-dominated dusty photoablating clouds.
This picture seems quite consistent with the present model.
In addition, it is reported that
Seyfert 1 galaxies have a smaller number of narrow-line clouds
than Seyfert 2s (Schmitt 1998), and also
the extension of NLRs in AGNs is roughly proportional to
the square root of the [O III] luminosity (Bennert et al. 2002). 
These trends are qualitatively compatible with the present results
that the number of dusty clouds decreases with a dimming
starburst, and the spatial extent of dusty clouds augments with
increasing AGN luminosity.

\section{CONCLUSIONS}

We have investigated the dynamics of gas clouds that
spurt out from circumnuclear starburst regions, including 
radiation forces from a starburst as well as an AGN.
The results are summarized as follows:

(1) In a case where the starburst luminosity is higher than
the AGN luminosity, gas clouds are distributed at several hundred pc
above the galactic disk, due to the radiation pressure from
the starburst.
The covering factor of these clouds reaches around 20\%
at the maximum. According as the starburst dims,
gas clouds with large optical depth are formed by 
cloud-cloud collisions and fall back due to the reduced radiation 
pressure by starburst. 
The larger gas clouds which have optical depth of several tens
are eventually distributed near to the equatorial plane.
This starburst-dominant case is qualitatively consistent with
the fact that Seyfert 2 galaxies appear to be more frequently
associated with starbursts than Seyfert 1s.

(2) In a case where the AGN luminosity is dominant and super-Eddington,
most gas clouds are blown out by the radiation pressure from the 
AGN. Resultantly, starburst-origin gas clouds can hardly 
contribute to the obscuration of the AGN. Hence, the AGN is likely
to be observed as type 1 with high probability. 
This AGN-dominant case corresponds to a quasar-like object.
Furthermore, it is predicted that 
the fraction of obscured AGNs decreases with increasing AGN
luminosity. This prediction matches the fraction of X-ray absorbed AGNs
found by Ueda et al. (2003).


(3) The velocity dispersion of gas clouds 
is several 100 km/s, which is just comparable to 
that found in NLRs. The observations for the ionization parameter of NLRs
and the extension of NLRs depending on AGN luminosity
are consistent with the present picture.

In the present analysis, we have found that starburst-origin gas 
clouds can contribute to the obscuration especially in large scales.
Also, it has turned out that there is still something missing
to account for all the properties of the AGN obscuration.
It may be significant to consider further effects such as 
the combination of inner and outer obscuring materials. 

\acknowledgments

\section*{ACKNOWLEDGMENTS}

We are grateful to N. Kawakatsu and K. Ohsuga for
helpful discussion. We also thank the anonymous referee for
valuable comments. This work was carried out with computational facilities 
at the Center for Computational Science, University of Tsukuba.
MU acknowledges Grants-in-Aid for Scientific Research
from MEXT, 16002003.

\newpage

\begin{figure*}[hbtp]
  \centering

 http://www.rccp.tsukuba.ac.jp/Astro/watabe/
	\caption{The distributions of gas clouds and
	optical depth to AGN. $L_{\rm{AGN}} =
	10^{10}L_{\odot}$ is assumed. We show snapshots
	at $10^{7}$, $2\times 10^{7}$, and $10^{8}$ yr. 
	The left panels display the cloud distributions
	projected on
	the $xy-$ and the $rz-$plane in the coordinates
	shown in Figure \ref{fig2}. The colors
	of points correspond to the radii of 
	gas clouds. The right panels show $\tau$ for the each direction
	$(\theta, \phi)$ using Mollweide's projection for the
	all-sky plot. The colors of points correspond to
	the levels of $\tau$.}
	\label{fig3}
\end{figure*}

\begin{figure*}[hbtp]
  \centering
	\includegraphics[width=10cm]{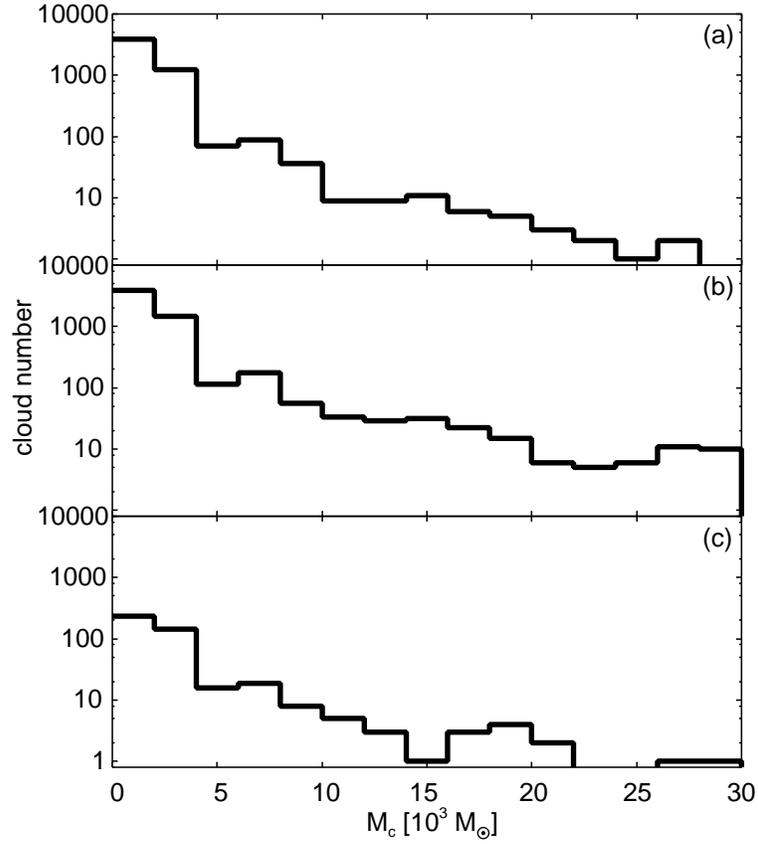}

	\caption{The mass spectrum of clouds at (a) $10^{7}$ yr, (b) $2\times
	10^{7}$ yr, and (c) $10^{8}$ yr. Horizontal axis is the 
	cloud mass in unit of
	$10^{3} M_{\odot}$, while vertical axis is the cloud
	number.}
	\label{fig4}
\end{figure*}

\begin{figure}[hbtp]
  \centering
	\includegraphics[width=11cm]{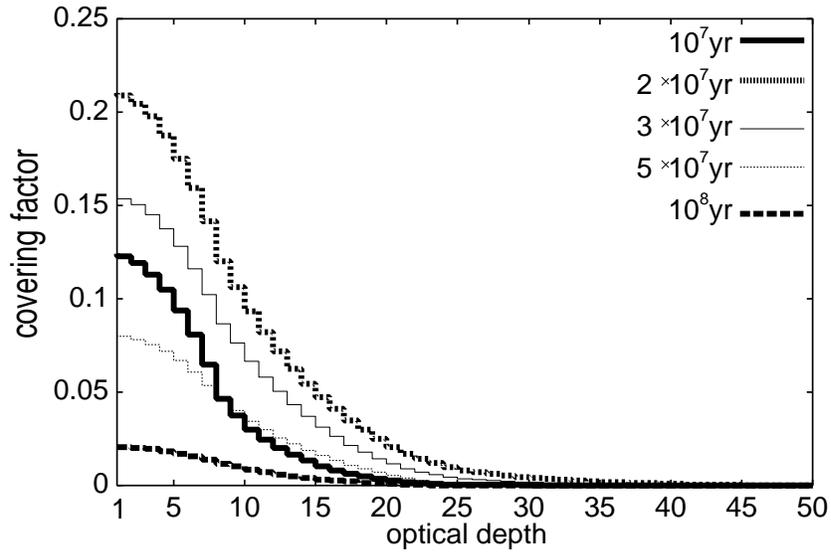}
	\caption{The covering factor of the area whose
	line-of-sight optical depth is greater than an
	optical depth given in the horizontal axis. $L_{\rm{AGN}} =
	10^{10}L_{\odot}$ is assumed. The
	evolution of the covering factor is shown at 1,
	2, 3, 5, and 10 $\times 10^{7}$ yr.}
	\label{fig5}
\end{figure}

\begin{figure}[hbtp]
  \centering
	\includegraphics[width=11cm]{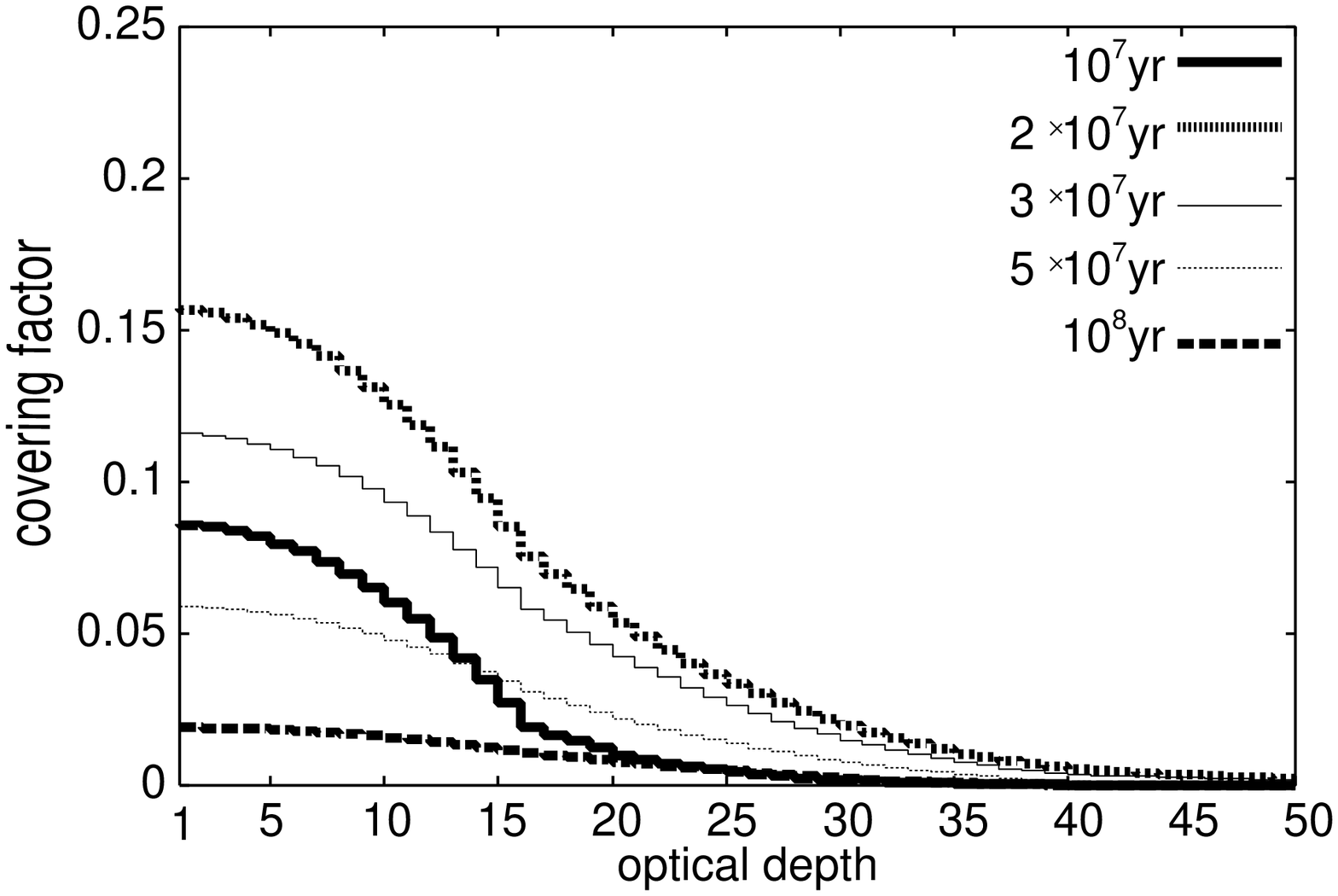}
	\includegraphics[width=11cm]{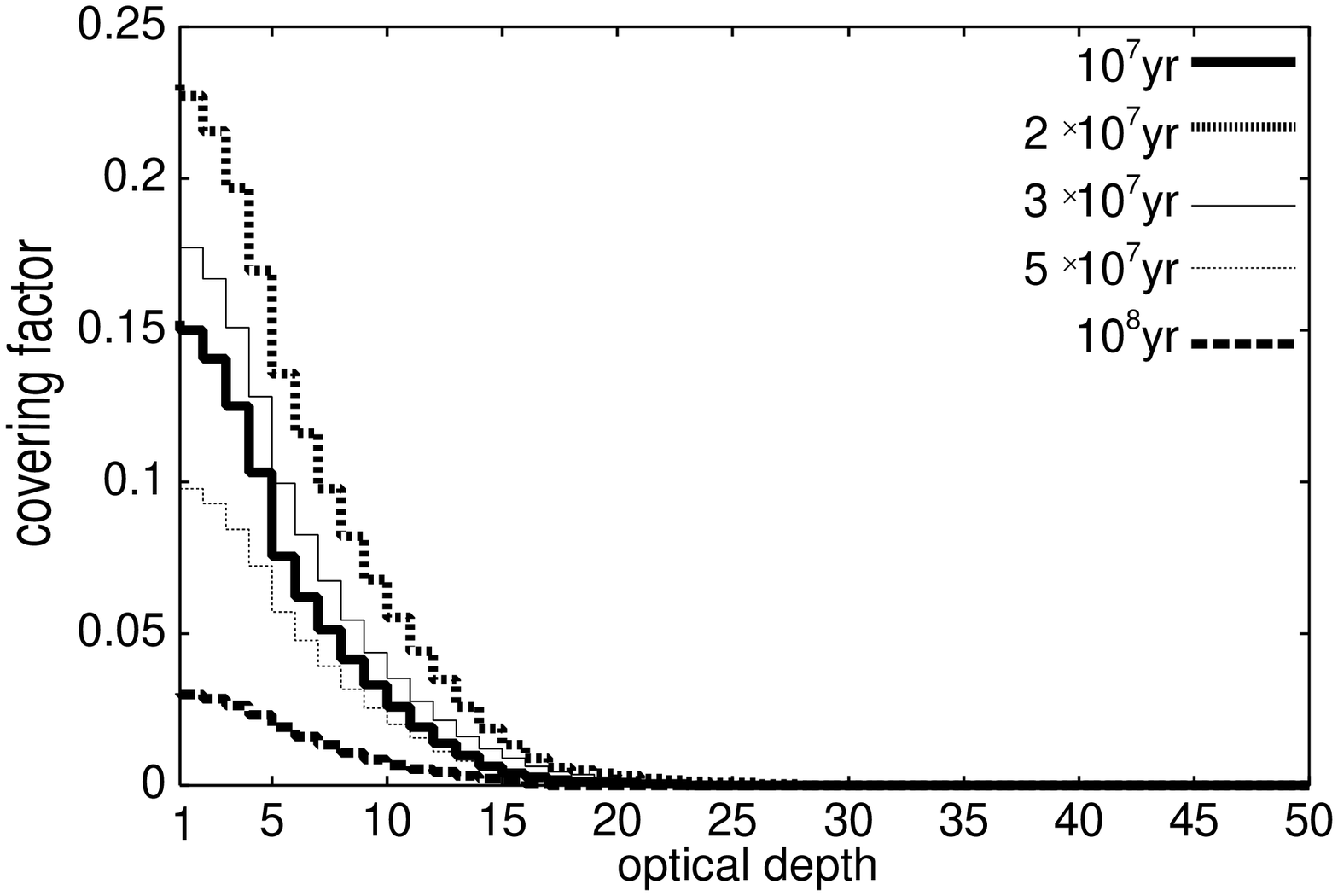}
	\caption{Same as Figure \ref{fig5}, but the initial
	cloud radius is 2.5 pc (above) and 4.5 pc (below).}
	\label{fig6}
\end{figure}

\begin{figure*}[hbtp]
  \centering
http://www.rccp.tsukuba.ac.jp/Astro/watabe/
	\caption{Same as Figure \ref{fig3}, but
	$L_{\rm{AGN}} = 3\times 10^{11}L_{\odot}$ is assumed. 
	In this case, since the
	radiation pressure from AGN is high, most of
	gas clouds cannot contribute to the
	obscuration of AGN.}
	\label{fig7}
\end{figure*}

\begin{figure}[hbtp]
  \centering
	\includegraphics[width=11cm]{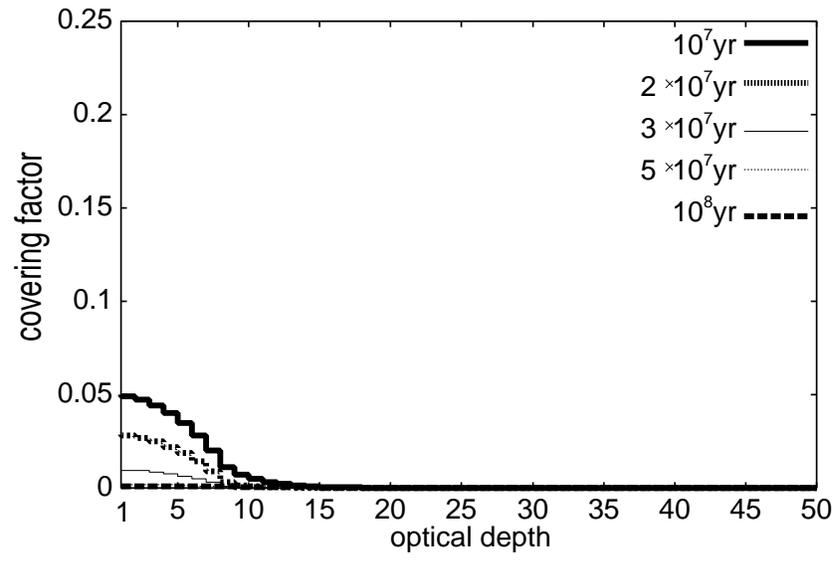}
	\caption{Same as Figure \ref{fig5}, but $L_{\rm{AGN}}
	= 3\times 10^{11}L_{\odot}$.}
	\label{fig8}
\end{figure}

\end{document}